\documentclass{article}

\usepackage{arxiv}

\usepackage[utf8]{inputenc} 
\usepackage[T1]{fontenc}    
\usepackage{hyperref}       
\usepackage{url}            
\usepackage{booktabs}       
\usepackage{amsfonts}       
\usepackage{nicefrac}       
\usepackage{microtype}      
\usepackage{lipsum}		
\usepackage{graphicx}
\usepackage{natbib}
\usepackage{doi}

\usepackage{amssymb}
\usepackage{amsmath}
\usepackage{bbm}

\usepackage[en-US]{datetime2}

\usepackage{graphicx}
\usepackage{caption}
\usepackage{subcaption}

\newcommand{\fbeta}{\tilde{\boldsymbol{\beta}}}
\newcommand{\fbetaMean}{\textbf{m}_{\tilde{\beta}}}
\newcommand{\fbetaSigma}{\Sigma_{\tilde{\beta}}}

\newcommand{\yv}{\textbf{y}}
\newcommand{\xm}{\tilde{\textbf{X}}}
\newcommand{\sigA}{a_{\sigma}}
\newcommand{\sigB}{b_{\sigma}}

\newcommand{\fbetaSigmaHat}{\hat{\Sigma}_{\tilde{\beta}}}
\newcommand{\fbetaMeanHat}{\hat{\textbf{m}}_{\tilde{\beta}}}
\newcommand{\sigAHat}{\hat{a}_{\sigma}}
\newcommand{\sigBHat}{\hat{b}_{\sigma}}

\newcommand{\etaAHat}{\hat{a}_{\eta}}
\newcommand{\etaBHat}{\hat{b}_{\eta}}

\title{Bayesian nonparametric scalar-on-image regression via Potts-Gibbs random partition models}

\author{
    {Mica Teo Shu Xian}\\
    School of Mathematics and Maxwell Institute for Mathematical Sciences \\
    University of Edinburgh \\
    James Clerk Maxwell Building \\
    Edinburgh, UK \\
    \texttt{mica.teo@ed.ac.uk} \\
	\And
	{Sara Wade}  \\
	School of Mathematics and Maxwell Institute for Mathematical Sciences \\
    University of Edinburgh \\
    James Clerk Maxwell Building \\
    Edinburgh, UK \\
	\texttt{sara.wade@ed.ac.uk} \\
}

\renewcommand{\shorttitle}{Potts-Gibbs SIR}

\hypersetup{
pdftitle={Potts-Gibbs SIR},
pdfsubject={q-bio.NC, q-bio.QM},
pdfauthor={Mica Teo Shu Xian, Sara Wade},
pdfkeywords={Bayesian nonparametric;  Gibbs-type priors; Potts model; Clustering; Generalised Swendsen-Wang; High-dimensional imaging data},
}

\begin{document}

\begin{center}
{\Large \textbf{Bayesian nonparametric scalar-on-image regression via Potts-Gibbs random partition models}} 
\\
\bigskip
Mica Teo Shu Xian\textsuperscript{1}
Sara Wade \textsuperscript{2},
\\
\bigskip
\textsuperscript{1,2}  School of Mathematics and Maxwell Institute for Mathematical Sciences, University of Edinburgh, United Kingdom
\\
\bigskip
\DTMlangsetup{showdayofmonth=true}

\today
\bigskip
\end{center}

\begin{abstract}
	Scalar-on-image regression aims to investigate changes in a scalar response of interest based on high-dimensional imaging data.  We propose a novel Bayesian nonparametric scalar-on-image regression model that utilises the spatial coordinates of the voxels to group voxels with similar effects on the response to have a  common coefficient.  We employ the Potts-Gibbs random partition model as the prior for the random partition in which the partition process is spatially dependent, thereby encouraging groups representing spatially contiguous regions. In addition, Bayesian shrinkage priors are utilised to identify the covariates and regions that are most relevant for the prediction. The proposed model is illustrated using the simulated data sets.
\end{abstract}


\section{Introduction}
\label{sec:introduction}

Through advances in data acquisition, vast amounts of high-dimensional imaging data are collected to study phenomena in many fields. Such data are common in biomedical studies to understand a disease or condition of interest \cite{CraddockR.Cameron2009Dspf, FanYong2008Safb, ShiJie2014Gioa,VanWalderveenM.A.A1998Hcoh}, and in other fields such as psychology \cite{DavatzikosChristos2005WMSo, SunDaqiang2009EaMR}, social sciences \cite{ferwerda2016using, HumNoelleJ2011Apiw, KimYunhwan2018Ucvt, SamanyNajmehNeysani2019Alef}, economics \cite{alma9923959053402466, NaikNikhil2016CAPT,  NaikNikhil2017Cvup}, climate sciences \cite{ONeillSaffronJ2013ImCc, ONeillSaffronJ2013Otuo}, environmental sciences \cite{DeboisDelphine2013MMIa, Gundlach-GrahamAlexander2015HHML, MaloofKatherineA2020Aoms} and more. While extracting features from the images based on predefined regions of interest favours interpretation and eases computational and statistical issues, changes may occur in only part of a region or span multiple structures.
In order to capture the complex spatial pattern of changes and improve accuracy and understanding of the underlying phenomenon, sophisticated approaches are required that utilize the entire high-dimensional imaging data.
However, the massive dimension of the images, which is often in the millions, combined with the relatively small sample size, which at best is usually in the hundreds, pose serious challenges. 

In the statistical literature, this is framed as a scalar-on-image regression (SIR) problem \cite{Ising-GMRF2014, Ising2013, Soft-thresholdedGP2018, Ising-DP2015}.
SIR belongs to the ``large p, small n" paradigm; thus, many SIR models utilise shrinkage methods that additionally incorporate the spatial information in the image \cite{Ising-GMRF2014, Ising2013, Soft-thresholdedGP2018, LeeKyoungjae2021Bgsi, Ising-DP2015,MehrotraSuchit2019SVSC, Reiss11,  SmithMichael2007SBVS,  wang2017generalized}.
In the SIR problem, the covariates represent the image value at a single pixel/voxel, i.e. a very tiny region, and the effect on the response is most often weak, unreliable and difficult to interpret. Moreover, neighbouring pixels/voxels are highly correlated, making standard regression methods, even with shrinkage, problematic due to multicollinearity.

To overcome these difficulties, we develop a novel Bayesian nonparametric (BNP) SIR model that extracts interpretable and reliable features from the images by grouping voxels with similar effects on the response to have a common coefficient. 
Specifically, we employ the Potts-Gibbs model \cite{LuHongliang2020Bnpf}   as the prior of the random image partition to encourage spatially dependent clustering. 
In this case, features represent regions that are automatically defined to be the most discriminative. This not only improves the signal and eases interpretability, but also reduces the computational burden by drastically decreasing the image dimension and addressing the multicollinearity problem. Moreover, it allows sharp discontinuities in the coefficient image across regions, which may be relevant in medical applications to capture irregularities \cite{wang2017generalized}.

In this direction, \cite{Ising-DP2015} proposed the Ising-DP SIR model, which combines an Ising prior to incorporate the spatial information in the sparsity structure with a Dirichlet Process (DP) prior to group coefficients. Still, the spatial information is only incorporated in the sparsity structure and not in the BNP clustering model, which could result in regions that are dispersed throughout the image. Instead, we propose to incorporate the spatial information in the random partition model,  encouraging spatially contiguous regions.  Further advantages of the nonparametric model include a data-driven number of clusters, interpretable parameters, and efficient computations. Moreover, we combine this with heavy-tailed shrinkage priors \cite{song2017}  to identify relevant covariates and regions.

The remainder of this article is organized as follows. Section \ref{sec:model} outlines the development of the SIR model based on the Potts-Gibbs models. Section \ref{sec:inference} derives the MCMC algorithm for posterior inference using the generalized Swendsen-Wang (GSW) \cite{daxu2016bayesian} algorithm for efficient split-merge moves that take advantage of the spatial structure. Section \ref{sec:simulation} illustrates the methods through simulation studies. Section \ref{sec:conclusion} concludes with a summary and future work.

\section{Model Specification}
\label{sec:model}
We introduce the statistical models that form the basis of the proposed Potts-Gibbs SIR model: SIR, random image partition model and shrinkage prior.

\subsection{Scalar-on-Image Regression}
SIR is a statistical linear method used to study and analyse the relationship between a scalar outcome and two or three-dimensional predictor images under a single regression model \cite{Ising-GMRF2014, Ising2013, Soft-thresholdedGP2018, Ising-DP2015}. For each data point, $i = 1, \ldots ,n$, we have
\begin{equation}
    y_i = \text{$\textbf{w}_i^T$}\boldsymbol{\mu} + \text{$\textbf{x}_i^T$} \boldsymbol{\beta} + \epsilon_i, \quad \text{$\epsilon_i \stackrel{iid}{\sim} \text{N}\left(0,\sigma^2\right),$}
\label{eq:SIregression}
\end{equation}
where $y_i$ is a  scalar continuous outcome measure, $\textbf{w}_i = (w_{i1}, \ldots ,w_{iq})^T \in \mathbb{R}^q$ is a $q$-dimensional vector of covariates, and $\textbf{x}_i= (x_{i1}, \ldots ,x_{ip})^T \in \mathbb{R}^p$ is a $p$-dimensional image predictor. Each $x_{ij}$ indicates the value of the image at a single pixel with spatial location $\textbf{s}_j = (s_{j1}, s_{j2})^T \in \mathbb{R}^2 $ for $j = 1, \ldots ,p$. We define $\boldsymbol{\mu} = (\mu_1, \ldots ,\mu_q)^T \in \mathbb{R}^q$ as a $q$-dimensional fixed effects vector and $\boldsymbol{\beta} = \left(\beta(\textbf{s}_1), \ldots ,\beta(\textbf{s}_p)\right)^T$ (with $\beta_j := \beta(\textbf{s}_j)$) as the spatially varying coefficient image described on the same lattice as $\textbf{x}_i$. We model the high-dimensional $\boldsymbol{\beta}$ by   spatially clustering the pixels into $M$ regions and assuming common coefficients $\beta^*_1, \ldots,  \beta^*_M$ within in each cluster, i.e. $\beta_j = \beta_m^*$ given the cluster label $z_j =m$. Thus, the prior on the coefficient image is decomposed into two parts: the random image partition model for spatially clustering the pixels and a shrinkage prior for the cluster-specific coefficients $\boldsymbol{\beta}^* = \left(\beta^*_1, \ldots,  \beta^*_M\right)^T$.    
The SIR model in \eqref{eq:SIregression} can be extended for other types of responses through a generalized linear model framework (GLM) \cite{mccullagh2019generalized}.

\subsection{Random Image Partition Model}
The image predictors are observed on a spatially structured coordinate system. Exchangeability is indeed no longer the proper assumption as the images contain covariate information, that we wish to leverage to improve model performance in this high-dimensional setting. 
To do so, we combine BNP random partition models, which avoid the need to prespecify the number of clusters, allowing it be determined and grow with the data, with  a Potts-like spatial smoothness component \cite{Potts_criticalValueFormula}. Spatial random partition models in this direction are a growing research area, including  Markov random field (MRF) with the product partition model (PPM) \cite{PanTianyu2020Ilgi}, with DP \cite{OrbanzPeter2008NBIS, daxu2016bayesian}, with Pitman–Yor process (PY)\cite{LuHongliang2020Bnpf} and with mixture of finite mixtures (MFM) \cite{HuGuanyu2020BSHP,ZhaoPeng2020BSHP}. Precisely, within the BNP framework, we focus on the class of  Gibbs-type random partitions \cite{Cerquetti2008GeneralizedCR, gnedin2006exchangeable, ModelsBeyondDP,CombinatorialStochasticProcesses}, motivated by their comprise between tractable predictive rules and richness of the predictive structure, including important cases, such as the DP \cite{ferguson1973bayesian}, PY \cite{perman1992size, JimPitman1996SDot}, and MFM \cite{miller2018mixture}. The Potts-Gibbs models induce a distribution on the partition $\pi_p = \{ C_1, \ldots, C_{M} \}$ of $p$ pixels into $M$ nonempty, mutually exclusive, and exhaustive subsets $C_1, \ldots, C_{M}$ such that $\cup_{C \in \pi_p} C = \{1, \ldots ,p\}$. The model can be summarised as:
    \begin{equation*}
    \text{pr}(\pi_p) \propto \exp \underbrace{\left(\sum_{j \sim k, j<k} \upsilon_{jk} \mathbbm{1}_{z_j=z_k} \right)}_{\text{Potts model}} \underbrace{\left( V_p(M) \prod^M_{m=1} W_m(\phi) \right)}_{\text{Gibbs-type random partition models}},
    \label{eq:PottsGibbsPartitionModel}
    \end{equation*}
where $z_j \in \{1, \cdots, M\}$, $j \sim k$ means that $j$ and $k$ are neighbors, and $\mathbbm{1}_{z_j=z_k}$ equals to 1 if $j$ and $k$ in the same cluster and 0 otherwise. In the following, we assume the spatial locations lie on a rectangular lattice with  first-order neighbors and a common coupling parameter $\upsilon$ for all neighbor pairs; a higher value of $\upsilon$ encourages more spatial smoothness in the partition.  We use the general notation $\phi$ to denote the parameters of the Gibbs-type partition models, and focus our study on three cases 1) DP with concentration parameter $\alpha > 0$; 2) PY with discount parameter $\delta \in [0,1)$ and concentration parameter $\alpha>-\delta$; and 3) MFM with parameter $\gamma>0$ (larger values encouraging more equally sized clusters) and a distribution $P_L(\cdot| \lambda)$ with parameter $\lambda$ related to the prior on the number of clusters. The $\{ V_p(M) : p \geq 1, 1 \leq M \leq p\}$ denotes the set of non-negative weights, which  solves the backward recurrence relation $V_p(M) = (p-\delta M) V_{p+1}(M) + V_{p+1}(M+1)$ with $V_1(1) = 1$. Table \ref{tab:Gibbscases} describes the $V_p(M)$ and $W_m(\phi)$ for DP, PY and MFM models.

    \begin{table}[!t]
        \caption{Formulas of $V_p(M), W_m(\phi)$ and terms of the predictive probability for assigning current cluster to either existing cluster or new cluster for DP, PY and MFM. Note that the predictive probabilities are stated up to a proportionality constant. }
        \label{tab:Gibbscases}
        \centering
        \begin{tabular}{c|c|c|c}
        & DP & PY & MFM \\ \hline 
            $ V_p(M)$ & $\frac{\Gamma( \alpha )\alpha^M}{\Gamma( \alpha+p)}$ & $ \frac{\Gamma(\alpha+1  ) \prod_{m=1}^{M-1} (\alpha + m\delta) }{\Gamma(\alpha + p )}$ & $ \sum_{l=1}^\infty  \frac{\Gamma(\gamma l) l!}{\Gamma(\gamma l + p) (l-m)!} P_L(\cdot| \lambda)$ \\
            $W_m(\phi)$ &  $\Gamma( \mid C_m \mid )$ & $\frac{\Gamma( \mid C_m \mid- \delta )}{\Gamma( 1- \delta) }$ & $\frac{\Gamma( \mid C_m \mid + \gamma )}{\Gamma( \gamma) }$ \\
            Existing cluster & $\frac{\Gamma(|C_{m}^{-A_o}| +|A_o| )}{\Gamma(|C_{m}^{-A_o}|)} $ & $\frac{\Gamma(|C_{m}^{-A_o}| +|A_o| - \delta)}{\Gamma(|C_{m}^{-A_o}|-\delta)}$ &  $\frac{\Gamma(|C_{m}^{-A_o}| +|A_o| + \gamma)}{\Gamma(|C_{m}^{-A_o}|+ \gamma)}$ \\
            New cluster & $ \alpha \Gamma(|A_o|)$ & $ (\alpha + \delta M^{-A_o}) \frac{\Gamma(|A_o| - \delta)}{\Gamma(1-\delta)}$ & $ \frac{V_{p}(M^{-A_o}+1)}{V_{p}(M^{-A_o})} \frac{\Gamma(|A_o| + \gamma)}{\Gamma(\gamma)}$
        \end{tabular}
    \end{table}

\subsection{Shrinkage Prior}

To identify relevant regions,  we use heavy tailed priors  for the unique values $(\beta_1^*, \ldots, \beta_M^*)$ of $\left( \beta(\textbf{s}_1), \ldots, \beta(\textbf{s}_p) \right)$.  Specifically, a $t$-shrinkage prior is used, motivated by its computational efficiency and nearly optimal contraction rate and selection consistency \cite{song2017}: 
\begin{equation}
\begin{aligned}
\sigma^{2} &\sim \text{IG}\left(a_{\sigma}, b_{\sigma}\right),\\ 
\left(\beta_m^*\right) | \sigma^{2} &\sim t_{\nu}(s \sigma), \quad \text{for all } m=1, \ldots, M,
\end{aligned}
\label{eq:t_shrinkage_prior}
\end{equation}
where $ t_{\nu}(s \sigma)$ denotes $t$-distribution with degree of freedom $\nu$ and scale parameter $s \sigma$. For posterior inference, 
the $t$-distribution \eqref{eq:t_shrinkage_prior} is rewritten as a hierarchical inverse-gamma scaled Gaussian mixture,
\begin{equation*}
\begin{aligned}
\sigma^{2} &\sim \text{IG}\left(a_{\sigma}, b_{\sigma}\right),\\
\eta_{m}^* &\sim \text{IG} \left(a_\eta, b_\eta \right), \\
\left(\beta_m^* \right) | \sigma^{2}, \eta_{m}^* &\sim N(0, \eta_{m}^* \sigma^{2}), \quad \text{for all } m=1, \ldots, M,
\end{aligned}
\label{eq:t_shrinkage_prior_2}
\end{equation*}
where $a_\eta $ and  $b_\eta$ are the shape and scaling parameter of the mixing distribution for each $\eta_m^*$ respectively with $\nu = 2 a_\eta$ and $s = \sqrt{b_\eta/a_\eta}$. 

\section{Inference}
\label{sec:inference}

We aim to infer the posterior distribution of the parameters based on the proposed Potts-Gibbs SIR model:

\begin{equation*}
    \begin{aligned}
    y_i \mid  \boldsymbol{\mu}, \boldsymbol{\beta}^*,  \pi_p, \sigma^2 &\sim \text{N}(\textbf{w}_i^T\boldsymbol{\mu} + \textbf{x}_i^{*T} \boldsymbol{\beta}^*, \sigma^2), \quad \text{for all $i=1, \ldots, n$},\\
    \boldsymbol{\mu} \mid \sigma^2 &\sim \text{N}(\textbf{m}_\mu, \sigma^2 \Sigma_\mu), \\
    \boldsymbol{\beta}^* \mid \boldsymbol{\eta}^*,  \sigma^2 &\sim \text{N}(\mathbf{0}_M, \sigma^2 \Sigma_{\beta^*}), \\
    \sigma^2 & \sim \text{IG}(a_{\sigma}, b_{\sigma}), \\
    \eta_{m}^* & \sim \text{IG} \left(a_\eta, b_\eta \right), \quad \text{for all $m=1, \ldots, M$}, \\
     \pi_p &\sim \text{Potts-Gibbs}(\upsilon, \phi),
    \end{aligned}
\label{eq:BNPFullModel}
\end{equation*}
where $x_{im}^* = \sum^p_{j=1} x_{ij} \mathbbm{1}(j \in C_m) / \sqrt{\mid C_m \mid}$ represents the total value, e.g. volume in the $m$th region of the image, $\textbf{m}_\mu = (m_{\mu_1} , \ldots , m_{\mu_q})$, $\Sigma_\mu = \text{diag}(c_{\mu_1}, \ldots, c_{\mu_q})^T$, and $\Sigma_{\beta^*} = \text{diag}(\eta_{1}^*, \ldots, \eta^*_{_M})$. Note that when defining $x_{im}^*$, we rescale by the square root of cluster size , which is equivalent to rescaling the variance of $\beta^*_m$  by the cluster size, encouraging more shrinkage for larger regions. 

We develop a Gibbs sampler to simulate from the posterior with a generalized Swendsen-Wang (GSW) algorithm 
to draw samples from the Potts-Gibbs model. 
Poor mixing can be seen in single-site Gibbs sampling \cite{GemanStuart1984SRGD} due to the high correlation between the pixel labels. 
The SW algorithm \cite{SwendsenRobertH1987Ncdi} addresses this by forming nested clusters of neighbouring pixels, then updating all of the labels within a nested cluster to the same value. 
The generalisation of the technique for standard Potts models to generalised Potts-partition models is called GSW \cite{daxu2016bayesian}. 
At each step of the algorithm, we proceed through the following steps:
\begin{enumerate}
    \item Sample the image partition $\pi_p$ given $\boldsymbol{\eta}^*$ and the data (with $\boldsymbol{\beta}^*, \boldsymbol{\mu}, \sigma^2$ marginalized). GSW is used to update simultaneously nested groups of pixels and hence improve the exploration of the posterior. The algorithm relies on the introduction of auxiliary binary bond variables, where $r_{jk} = 1$ if pixels $j$ and $k$ are bonded, otherwise 0.
    The bond variables define a partition of the pixels into nested clusters $A_1, \ldots , A_O$, where $O$ denotes the number of nested clusters and each $A_o \subseteq C_m$ for some $m=1, \ldots, M$. 
 For each neighbor pair $j \sim k$ for $1 \leq j < k \leq p$, we sample the bond variables as follows, $r_{jk} \sim \text{Ber}\{ 1 -  \exp(- \upsilon_{jk} \zeta_{jk} \mathbbm{1}_{z_j=z_k})\}$,
where we define $\zeta_{jk} = \kappa \exp\{-\tau d(\hat{\beta}_{j}, \hat{\beta}_{k})\}$ with  $\hat{\beta}_{j}$ denoting the estimated coefficient from univariate regression on the $j$th pixel and $\kappa, \tau$ are the tuning parameters of the GSW sampler. Notice that the algorithm reduces to single-site Gibbs when $\kappa=0$, 
and recovers classical SW when $\kappa=1$ and $\tau =0$. 

As we are dealing with non-conjugate priors, we update the cluster assignment by extending Gibbs sampling with the addition of auxiliary parameters, which is widely known as Algorithm 8 \cite{NealRadfordM2000MCSM}. 
We denote by $A_o$ the current nested cluster; $C_1^{-A_o}, \ldots ,C_M^{-A_o}$ the clusters without nested cluster $A_o$;  $M^{-A_o}$ the number of distinct clusters excluding $A_o$ and $h$ the number of temporary auxiliary variables.
For each nested cluster $A_o$, it is assigned to an existing cluster $m=1, \ldots ,M^{-A_o}$ or a new cluster $m = M^{-A_o}+1, \ldots ,M^{-A_o}+h$ with probability as follows, 
    \begin{equation*}
    \begin{aligned}
    &\text{pr}(A_o \in C_m^{-A_o} \mid  \cdots) \\
    &\propto \begin{cases}
      \frac{\Gamma(|C_{m}^{-A_o}| +|A_o| - \delta)}{\Gamma(|C_{m}^{-A_o}|-\delta)}  \text{pr}\left( \textbf{y}  \mid  \pi_p^{A_o \rightarrow m} , \boldsymbol{\eta^*}\right) \\
     \prod_{ \{ (j,k)  \mid  j \in A_o, k \in C_m^{-A_o}, r_{jk}=0 \} } \exp \left\{ \upsilon_{jk} (1-\zeta_{jk})  \right\} , & \text{for } C_m^{-A_o} \in \pi_p^{-A_o},\\ \\
    \frac{1}{h}\frac{V_{p}(M^{-A_o}+1)}{V_{p}(M^{-A_o})} \frac{\Gamma(|A_o| - \delta)}{\Gamma(1-\delta)} \text{pr}\left( \textbf{y}  \mid  \pi_p^{A_o \rightarrow M+1} , \boldsymbol{\eta^*} \right) ,  &  \text{for new $C_m^{-A_o}$};
    \end{cases}
    \end{aligned}
    \label{eq:PottsGibbssampler}
    \end{equation*}
where $\text{pr}\left( \textbf{y}  \mid  \pi_p^{A_o \rightarrow m} , \boldsymbol{\eta^*} \right)$ and $\text{pr}\left( \textbf{y}  \mid  \pi_p^{A_o \rightarrow M+1} , \boldsymbol{\eta^*} \right)$ denote the marginal likelihood of data obtained by moving $A_o$ from its current cluster to existing clusters or newly created cluster respectively.
Before updating the cluster assignments, we sample the nested clusters and compute the volume of each nested cluster for all images, with computational cost $\mathcal{O}( np)$. When updating the cluster assignments,  the marginal likelihood dominates the computational cost, as it involves inversion and determinants of $(M+q)\times (M+q)$ matrices and updating the sufficient statistics for every nested cluster and every outer cluster allocation, i.e. the cost is $\mathcal{O}([[M+q]^3 + n[M+q]] OM )$. 

\item Sample $\boldsymbol{\beta}^*, \boldsymbol{\mu}, \sigma^2$ jointly given  the partition $\pi_p$, $\boldsymbol{\eta}^*$ and the data.
    Notationally, we reformulate $\tilde{\textbf{x}}_i = ( \textbf{w}_i^T, \textbf{x}^{*\, T}_i )^T$ and $\fbeta = (\boldsymbol{\mu}^T, \boldsymbol{\beta}^{* \, T} )^T$.
We define $\xm $ be the matrix with rows equal to $\tilde{\textbf{x}}_i^T$. 
The corresponding full conditional for $\fbeta$ and $\sigma^2$ is
\begin{equation*}
\begin{aligned}
\sigma^2 \mid \cdots \sim& \text{IG} ( \sigAHat, \sigBHat ),\\
\fbeta \mid \sigma^2, \cdots \sim& \text{N} ( \fbetaMeanHat, \sigma^2\fbetaSigmaHat), 
\end{aligned}    
\end{equation*}
where $\fbetaSigmaHat = ( \fbetaSigma^{-1} +\xm^T \xm) ^{-1}$, $\fbetaMeanHat = \fbetaSigmaHat ( \fbetaSigma^{-1} \fbetaMean + \xm^T \yv)$, and  $\text{IG}(\sigAHat, \sigBHat)$ denotes the inverse-gamma distribution with updated shape $\sigAHat = \sigA + n/2$ and scale $\sigBHat = \sigB + [ \fbetaMean^T\fbetaSigma^{-1}\fbetaMean+\yv^T\yv -  \fbetaMeanHat^T\fbetaSigmaHat^{-1}\fbetaMeanHat]/2$.

\item Sample $\boldsymbol{\eta}^*$ given $\boldsymbol{\beta}^*$.
The corresponding full conditional for each $\eta_m^*$ is an inverse-gamma distribution with updated shape $\etaAHat = a_\eta+1/2$ and scale $\etaBHat = b_\eta + (\beta^*_m)^2/(2\sigma^2)$: 
\begin{equation*}
\begin{aligned}
\eta_m^* \mid \cdots \sim \text{IG} (\etaAHat, \etaBHat), \quad \text{for  $m=1, \ldots, M$}.
\end{aligned}
\end{equation*}
\end{enumerate}

\section{Numerical Studies}
\label{sec:simulation}

\begin{figure}[tb!]
\begin{center}
\begin{subfigure}[b]{0.33\textwidth}
\centering\includegraphics[trim=0cm 0cm 0cm 0cm, clip, width=1.1\textwidth]{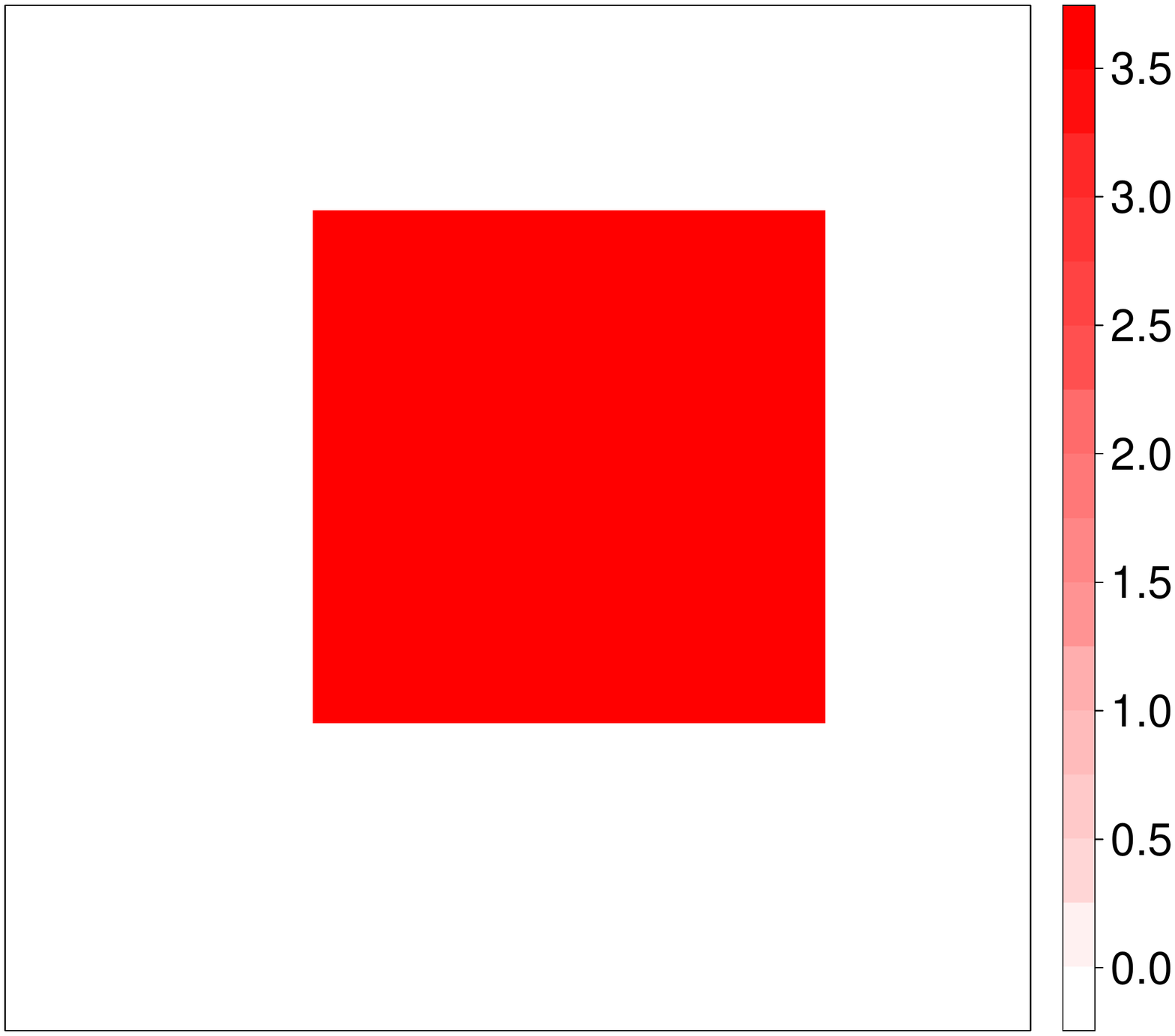}
\caption{Scenario 1: Truth}
\end{subfigure}%
\begin{subfigure}[b]{0.33\textwidth}
 \centering\includegraphics[trim=0cm 0cm 0cm 3cm, clip, width=1.2\textwidth]{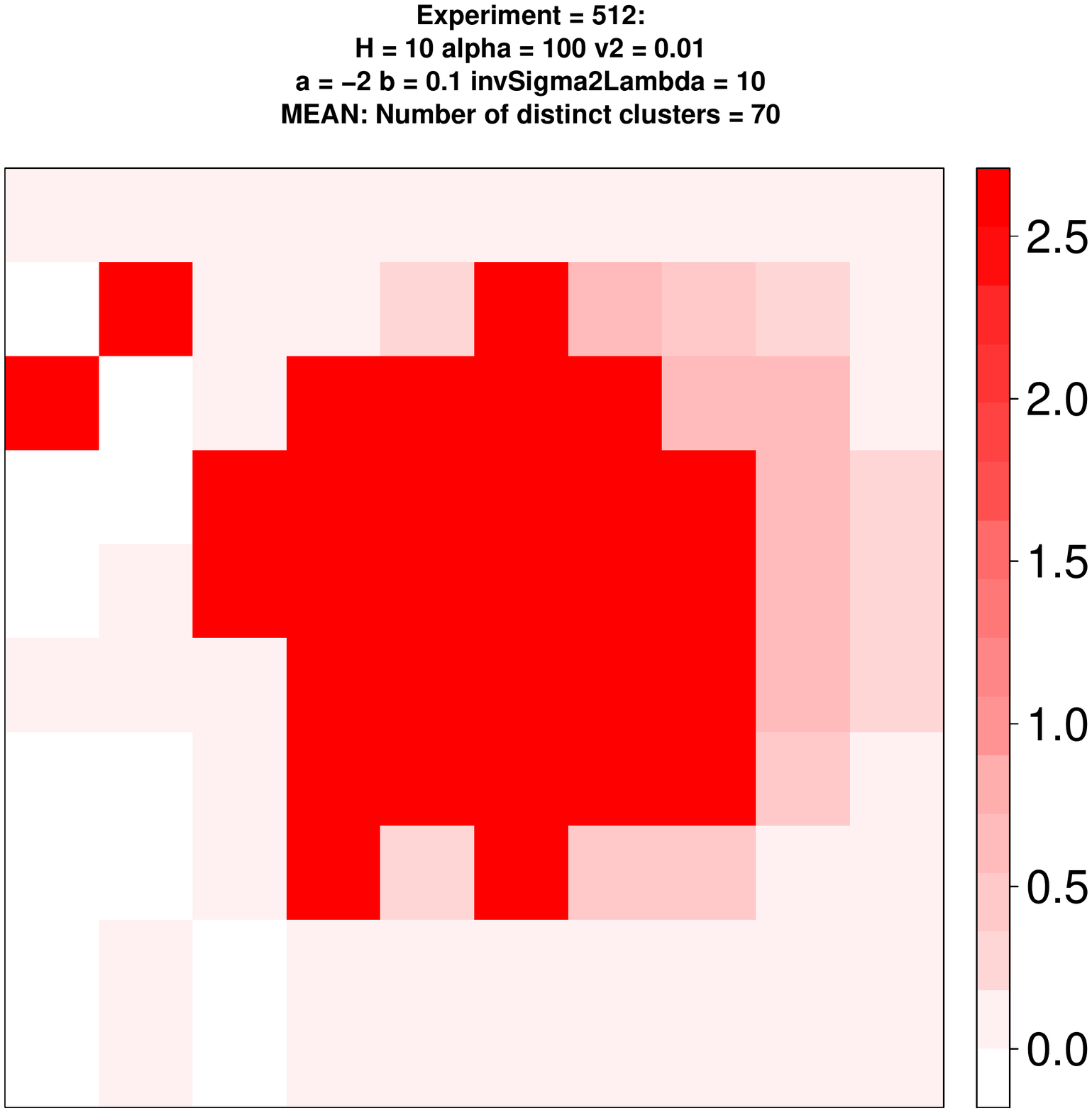}
\caption{Scenario 1: Ising-DP}
\end{subfigure}%

\begin{subfigure}[b]{0.33\textwidth}
\centering\includegraphics[trim=0cm 0cm 0cm 3cm, clip, width=1.2\textwidth]{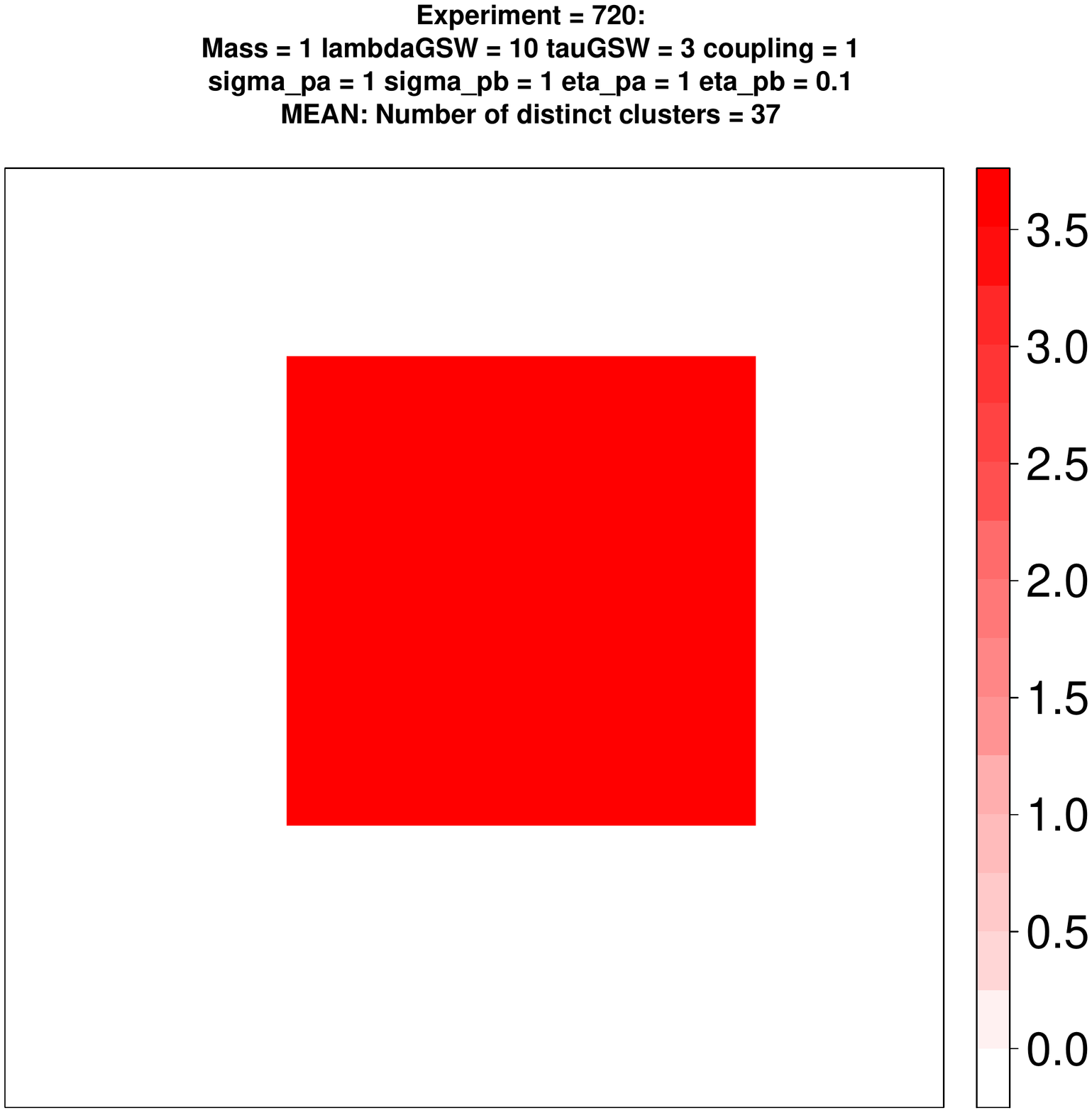}
\caption{Scenario 1: Potts-DP}
\end{subfigure}%
\begin{subfigure}[b]{0.33\textwidth}
\centering\includegraphics[trim=0cm 0cm 0cm 3cm, clip, width=1.2\textwidth]{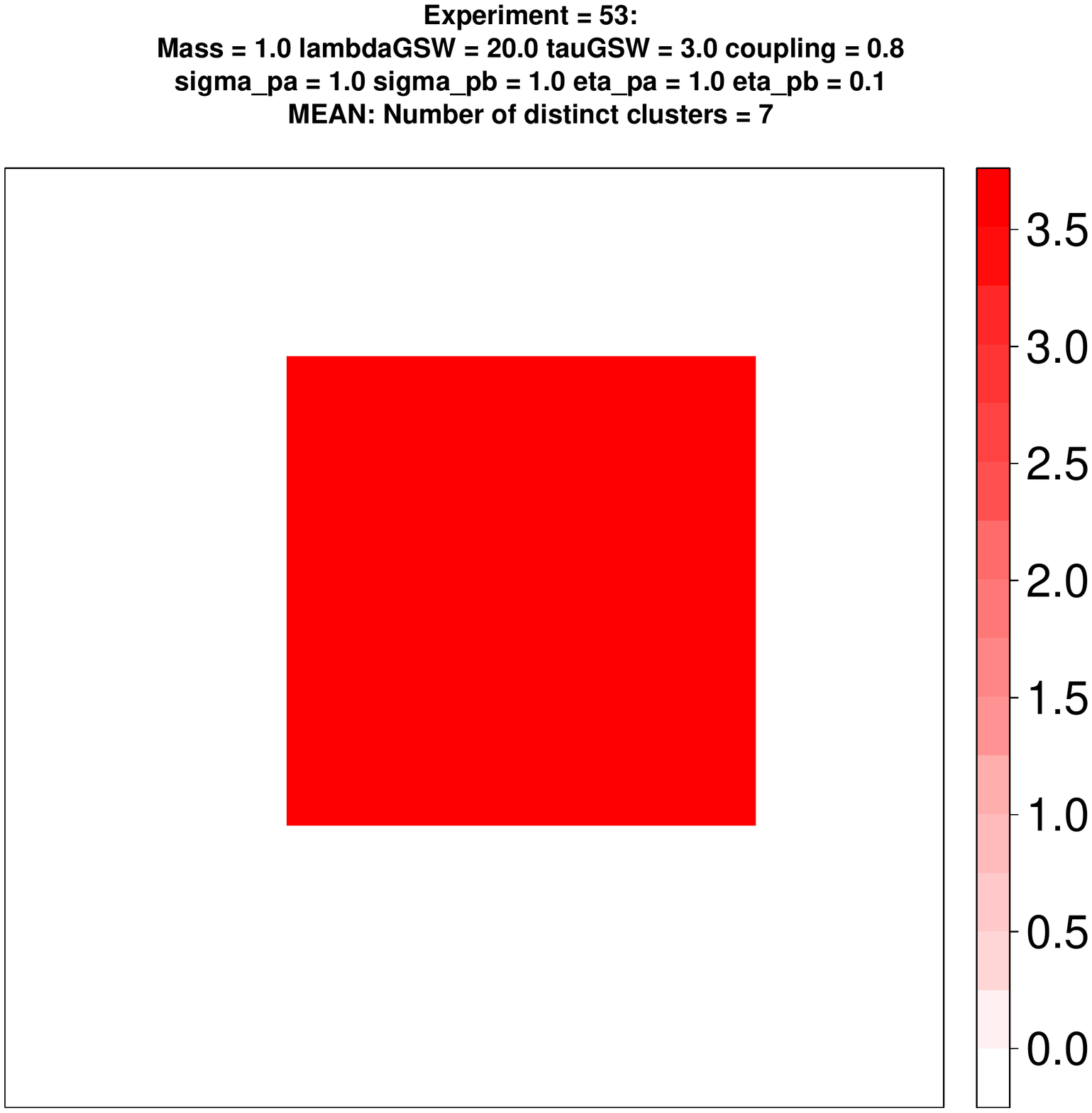}
\caption{Scenario 1: Potts-PY}
\end{subfigure}%
\begin{subfigure}[b]{0.33\textwidth}
\centering\includegraphics[trim=0cm 0cm 0cm 3cm, clip, width=1.2\textwidth]{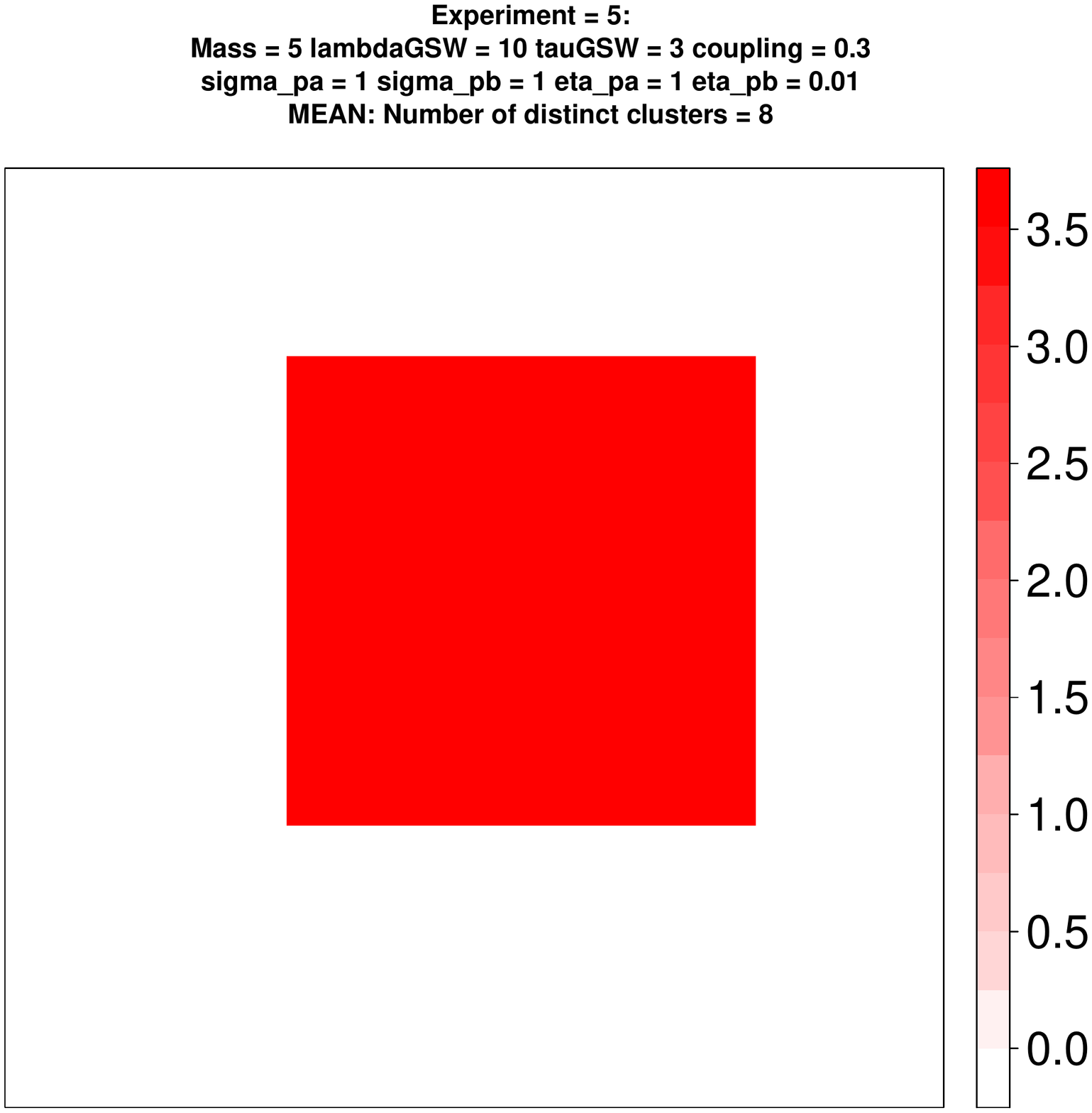}
\caption{Scenario 1: Potts-MFM}
\end{subfigure}%
\end{center}
\caption{Figures on the upper and bottom row showing the true and estimated coefficient matrix of the simulated data sets for scenario 1 under each model.}
\label{fig:truebestResultSim1}
\end{figure}

\begin{figure}[tb!]
\begin{center}
\begin{subfigure}[b]{0.33\textwidth}
\centering\includegraphics[trim=0cm 0cm 0cm 0cm, clip, width=1.1\textwidth]{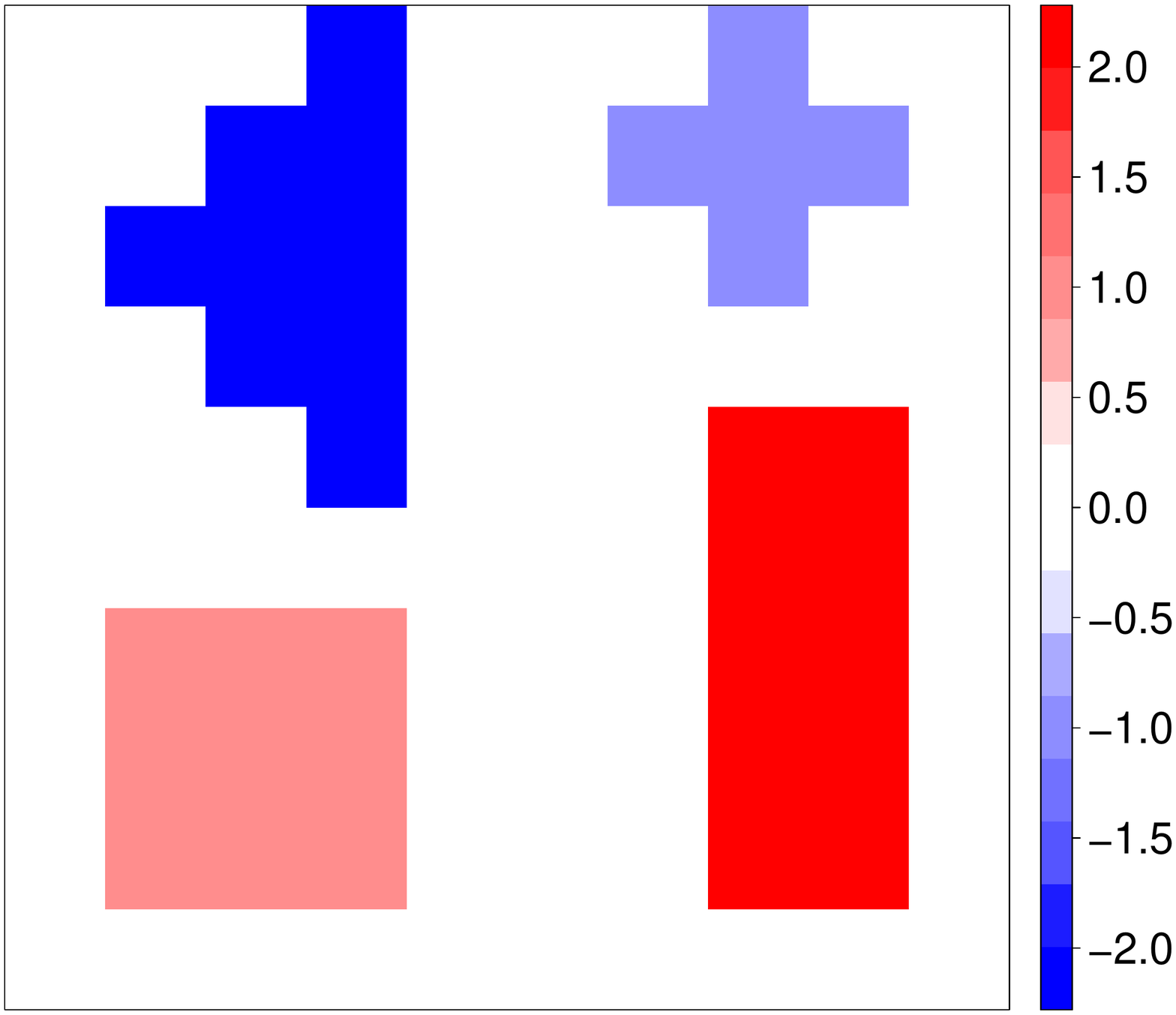}
\caption{Scenario 2: Truth}
\end{subfigure}%
\begin{subfigure}[b]{0.33\textwidth}
 \centering\includegraphics[trim=0cm 0cm 0cm 3cm, clip, width=1.2\textwidth]{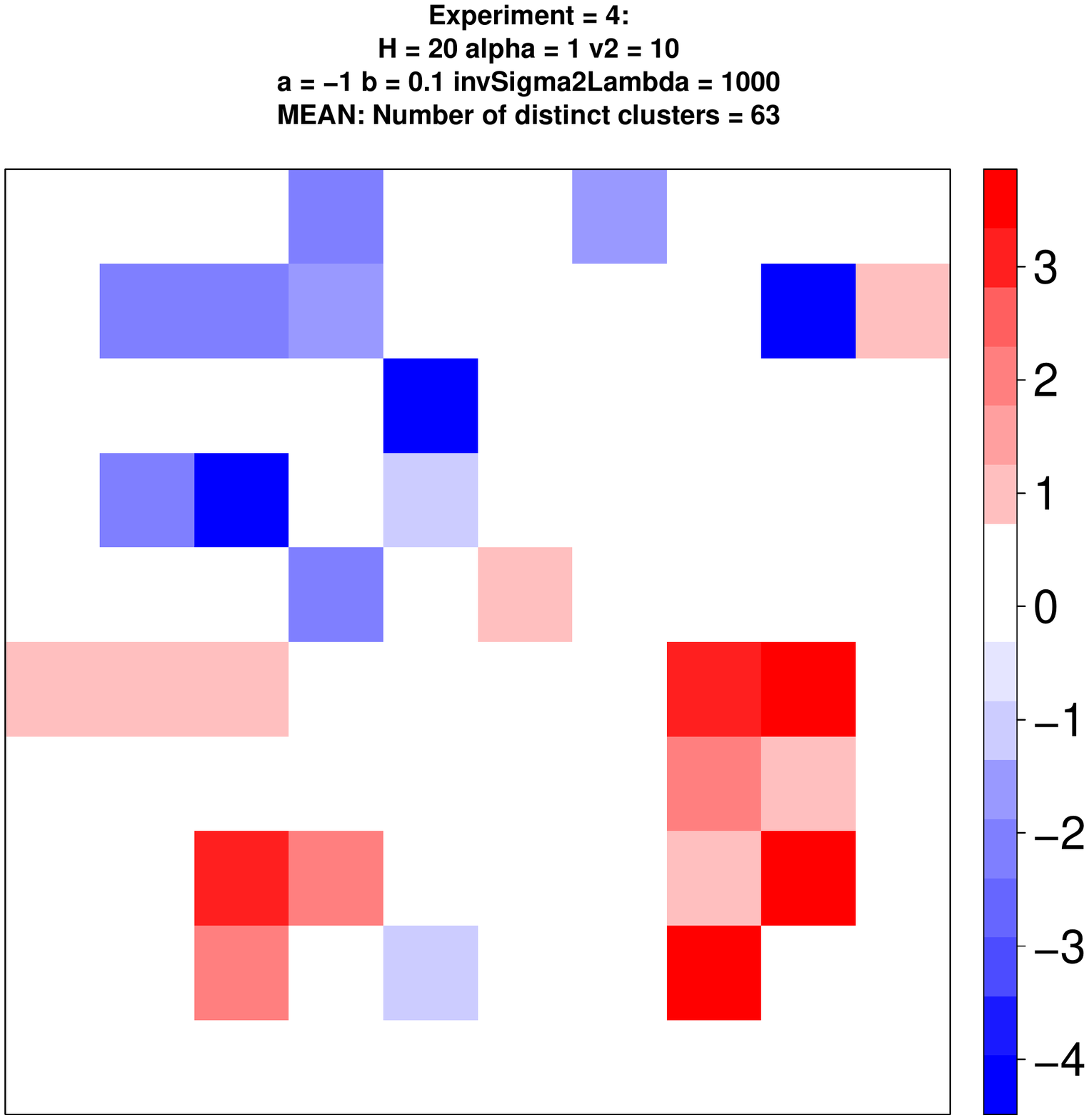}
\caption{Scenario 2: Ising-DP}
\end{subfigure}%

\begin{subfigure}[b]{0.33\textwidth}
\centering\includegraphics[trim=0cm 0cm 0cm 3cm, clip, width=1.2\textwidth]{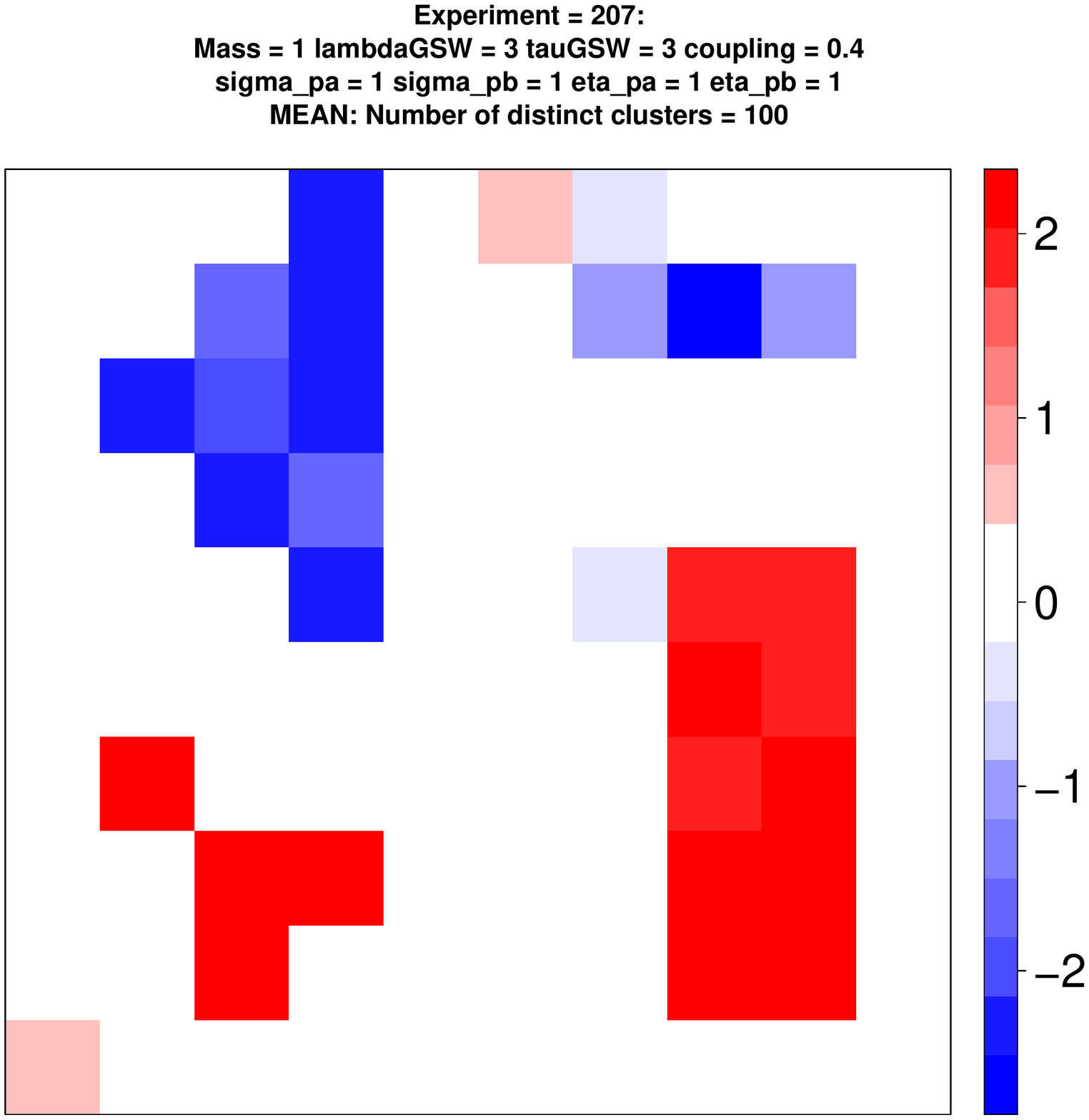}
\caption{Scenario 2: Potts-DP}
\end{subfigure}%
\begin{subfigure}[b]{0.33\textwidth}
\centering\includegraphics[trim=0cm 0cm 0cm 3cm, clip, width=1.2\textwidth]{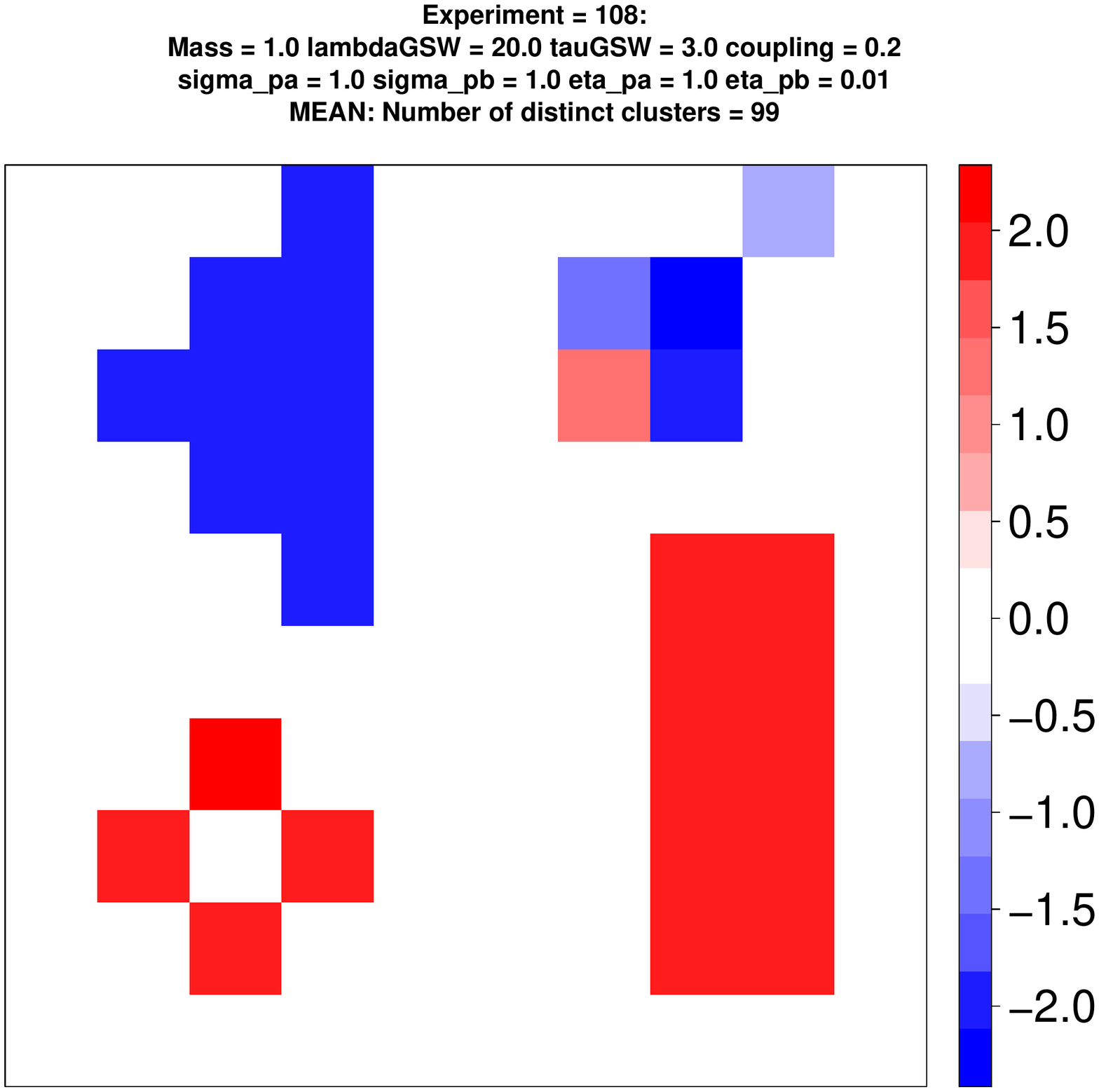}
\caption{Scenario 2: Potts-PY}
\end{subfigure}%
\begin{subfigure}[b]{0.33\textwidth}
\centering\includegraphics[trim=0cm 0cm 0cm 3cm, clip, width=1.2\textwidth]{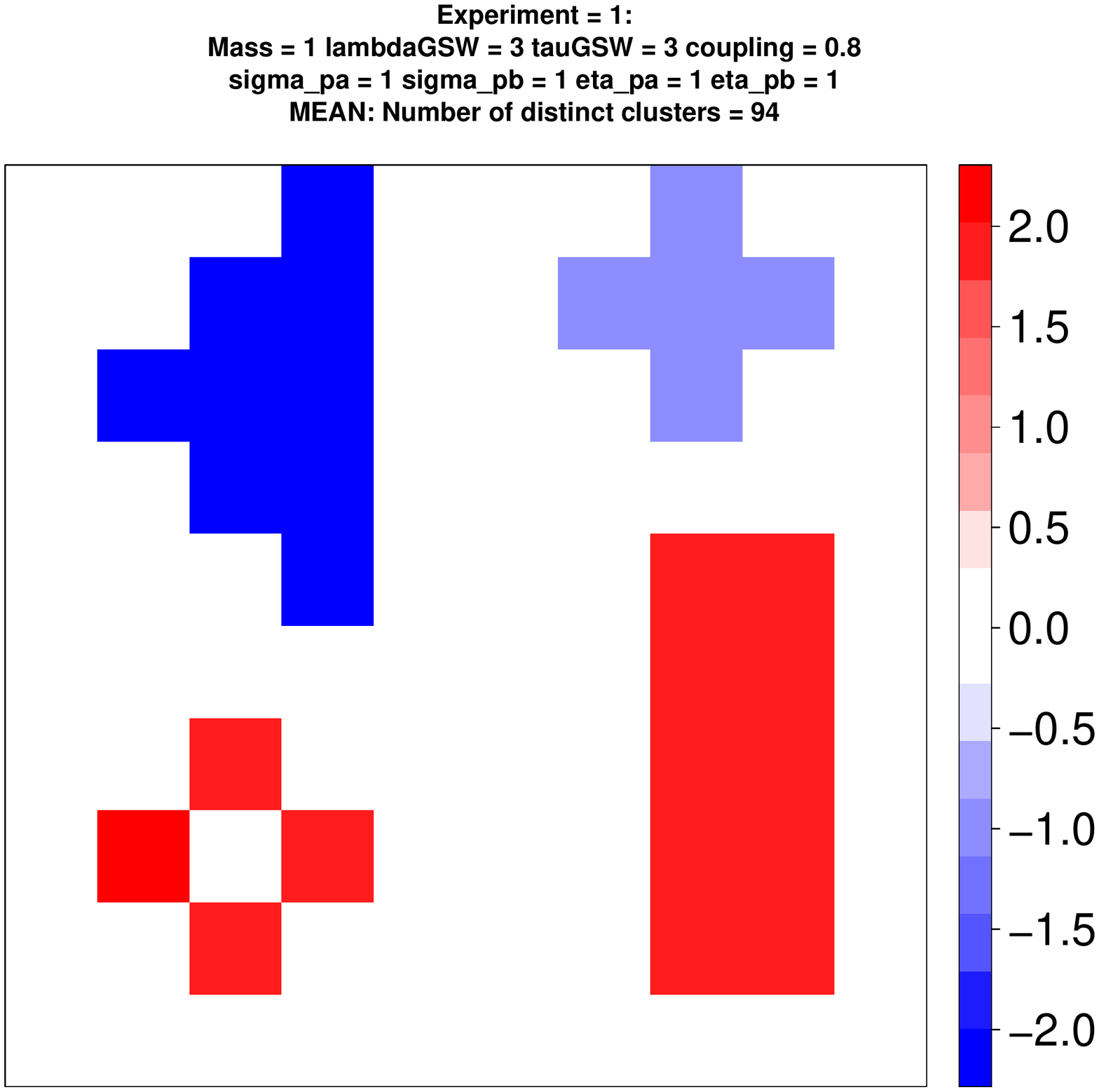}
\caption{Scenario 2: Potts-MFM}
\end{subfigure}%
\end{center}
\caption{Figures on the upper and bottom row showing the true and estimated coefficient matrix of the simulated data sets for scenario 2 under each model.}
\label{fig:truebestResultSim2}
\end{figure}

We study through simulations the performance of the proposed model and compare it with Ising-DP \cite{Ising-DP2015}. 
We consider 2D images in this simulation. 
The $n=300$ images are simulated on a two dimensional grid of size $10 \times 10$, with spatial locations $\textbf{s}_j = (s_{j1}, s_{j2}) \in \mathbf{R}^2$ for $1\leq s_{j1}, s_{j2} \leq 10$. For simplicity's sake, we include an intercept but do not consider others covariates, $\textbf{w}_i$. 
We concentrate on the two simulation scenarios with true $M=2$ and $M=5$ as shown in Figures \ref{fig:truebestResultSim1} - \ref{fig:truebestResultSim2}. For each experiment, we summarise the posterior of the clustering structure of the data sets by minimising the posterior expected Variation of Information (VI) \cite{minVIWade_2018}.

The Potts-Gibbs models can detect correctly the cluster structure under scenario 1 (Figure \ref{fig:truebestResultSim1}). The Potts-Gibbs models are also capable of capturing and identifying the more complex cluster structure underlying the data for scenario 2 (Figure \ref{fig:truebestResultSim2})  with the ARI 0.621 - 0.830 (Table \ref{tab:resultSimData2}). On the contrary, Ising-DP has failed terribly to recover the cluster structure for scenario 2, as illustrated in Figure \ref{fig:truebestResultSim2}.  It is observed that under the Potts-Gibbs models, most of the resultant clusters are spatially proximal, while under Ising-DP, the clusters are dispersed throughout the image. 
By taking into consideration spatial dependence in the random partition model via the Potts-Gibbs models, the proposed models produce spatially aware clustering and thus improve the predictions. 

DP has a concentration parameter $\alpha$, with larger values encouraging more new clusters and a rich-get-richer property that favours allocation to larger clusters. The PY has an additional discount parameter $\delta \in [0,1)$ that helps to mitigate the rich-get-richer property and phase transition of the Potts model. The MFM has a parameter $\gamma$, with larger values encouraging more equal-sized clusters and helping to avoid phase transition of the Potts model, as well as additional parameters $\lambda$ which are related to the prior on the number of clusters.

\begin{table}[tb!]
\centering
\resizebox{\textwidth}{!}{%
\begin{tabular}{@{}llllll@{}}
\toprule
\multicolumn{6}{c}{Scenario 1} \\
 & ARI & VI & MSE & MSPE & $M$ \\ \midrule
Potts-DP & \textbf{1.0} (0.004) & 0.001 (0.010) & 1.33e-4 (5.59e-4) & 4.215 (0.057) & 2.019 (0.138) \\
Potts-PY & \textbf{1.0} (0.004) & 0.001 (0.009) & 1.03e-4 (8.73e-5) & 4.213 (0.052) & 2.015 (0.122) \\
Potts-MFM & \textbf{0.999} (0.007) & 0.001 (0.014) & 1.01e-4 (8.37e-5) & 4.209 (0.052) & 2.007 (0.081) \\
Ising-DP & 0.307 (0.079) & 1.386 (0.154) & 0.807 (0.011) & 145.912 (10.051) & 4.575 (1.340) \\ \midrule \hline
\multicolumn{6}{c}{Scenario 2} \\
 & ARI & VI & MSE & MSPE & $M$ \\ \midrule
Potts-DP & \textbf{0.621}(0.060) & 1.160 (0.211) & 0.246 (0.064) & 7.754 (2.653) & 6.722 (0.901) \\
Potts-PY & \textbf{0.713} (0.050) & 1.006 (0.147) & 0.157 (0.035) & 0.868 (0.168) & 6.882 (1.090) \\
Potts-MFM & \textbf{0.830} (0.036) & 0.599 (0.133) & 0.093 (0.014) & 0.850 (0.122) & 5.232 (0.475) \\
Ising-DP & 0.038 ( 0.021) & 3.990 (0.159) & 0.980 ( 0.025) & 3.641 (0.526) & 15.542 (1.554) \\ \bottomrule
\end{tabular}%
}
\vspace*{0.1cm}
\caption{Mean and standard deviation of the posterior of adjusted Rand index (ARI), variation information (VI), mean squared error  (MSE), mean squared prediction error (MSPE), and number of clusters for each scenario under each model.}
\label{tab:resultSimData2}
\end{table}

\section{Conclusion}
\label{sec:conclusion}
We have developed novel Bayesian scalar-on-image regression models to extract interpretable features from the image by clustering and leveraging the spatial coordinates of the pixels/voxels.  To encourage groups representing spatially contiguous regions, we incorporate the spatial information directly in the prior for the random partition through  Potts-Gibbs random partition models. 
We have shown the potential of  Potts-Gibbs models in detecting the correct cluster structure on simulated data sets. In our experiments, the hyperparameters of the Potts-Gibbs model were determined via a simple grid search on selected combinations of hyperparameters. 
However, future work will consist of investigating the influence of the various parameters inherent to the model and guidelines and tools to determine hyperparameters. The model will then be applied to real images, e.g. neuroimages. Motivated by examining and identifying brain regions of interest in Alzheimer's disease, we will use  MRI images obtained from the Alzheimer's Disease Neuroimaging Initiative (ADNI) database   \url{(www.adni-info.org)}. The proposed SIR model will be extended to classification problems through the GLM framework.

\bibliographystyle{unsrtnat}
\bibliography{main}  






\end{document}